\documentclass[article]{aa}

\usepackage{natbib}
\usepackage{graphicx}
\usepackage{txfonts}

\begin{document}

\title{The effect of longitudinal flow on resonantly damped kink oscillations }
\author{J. Terradas$^{1}$, M. Goossens$^{1}$, I. Ballai$^{2}$}

\offprints{J. Terradas, \email{jaume@wis.kuleuven.be}}
\institute{$^1$Centre Plasma Astrophysics and Leuven Mathematical Modeling and
Computational Science Center, Katholieke Universiteit Leuven, Celestijnenlaan
200B, B-3001 Leuven, Belgium, \email{jaume@wis.kuleuven.be, marcel.goossens@wis.kuleuven.be}
\\$^2$Solar Physics and Space Plasmas Research
Centre (SP$^2$RC), Department of Applied Mathematics, University of Sheffield, Hounsfield Road, Hicks Building, Sheffield, S3 7RH, England, UK\email{I.Ballai@sheffield.ac.uk}
}
{}

\date{Received / Accepted }

\abstract{The most promising mechanism acting towards damping the kink
oscillations of coronal loops is resonant absorption. In this context most of
previous studies neglected the effect of the obvious equilibrium flow along
magnetic field lines. The flows are in general sub-Alfv\'enic and hence
comparatively slow.}{ Here we investigate the effect of an equilibrium flow on
the resonant absorption of linear kink MHD waves in a cylindrical magnetic flux
tube with the aim of determining the changes in the frequency of the forward and
backward propagating waves and in the modification of the damping times due to
the flow.}{A loop model with both the density and the longitudinal flow changing
in the radial direction is considered. We use the thin tube thin boundary (TTTB)
approximation in order to calculate the damping rates. The full resistive
eigenvalue problem is also solved without assuming the TTTB
approximation.}{Using the small ratio of flow and Alfv\'en speeds we derive
simple analytical expressions to the damping rate. The analytical expressions
are in good agreement with the resistive eigenmode calculations.}{Under typical
coronal conditions the effect of the flow on the damped kink oscillations is
small when the characteristic scale of the density layer is similar or smaller
than the characteristic width of the velocity layer. However, in the opposite
situation the damping rates can be significantly altered, specially for the
backward propagating wave which is undamped while the forward wave is
overdamped.}


\titlerunning{Resonantly damped kink oscillations with flow}
\authorrunning{Terradas et al.}
\maketitle

\section{Introduction}

Dynamism is probably one of the most fundamental characteristics of space and
solar plasmas, flows of plasma being observed in high resolution on almost all
temporal and spatial scales. Flows are ubiquitous in active region loops and the
measurements of their velocities have been provided by instruments like SoHO
\citep[see][]{brekkeetal97,winetal02}, TRACE \citep[see][]{winetal01} and more
recently Hinode \citep[see for example][]{chaeetal08,ofmwang08,terrarr08}. In
general the flow speeds are small, and in most of the observations they are
sub-Alfv\'enic, typically less that 10\% of the Alfv\'en speed. The bulk motions
are observed along magnetic field lines which outline coronal loops. These flows
could be generated by some catastrophic cooling  of coronal loops or are related
to some siphon mechanism arising due to the difference in pressure at the loop
footpoints. Since longitudinal steady flows carrying momentum and providing
additional inertia are present in coronal loops it is necessary to study  their
effects on the transverse oscillations observed in these structures \citep[see
for example][]{aschetal99,nakaetal99}. An effect which is of obvious importance
for magnetohydrodynamic (MHD) wave theory of loops' dynamics is how the period
and the damping time are modified by the flow.

Here we are interested in the damping of the fundamental kink mode due to
resonant absorption, based on the transfer of energy from a global MHD wave to
local resonant Alfv\'{e}n waves, and how the efficiency of the mechanism is
altered by a stationary flow. In the past, the influence of a velocity shear on
this process due to a longitudinal flow has been studied by
 \citet{holletal90,peredotata90},
\citet{rudgoss95,erdetal95,erdgooss96,tirryetal98}. More recently,
\citet{andriesetal00,andgooss01,erdtaro03a,erdtaro03}, have investigated in
detail resonant flow instabilities which can occur for velocity shears
significantly below the Kelvin-Helmholtz (KH) threshold. These instabilities are
produced when the frequency of the forward propagating wave (propagating in the
direction of the flow) shifts into the Doppler shifted continuum of the backward
propagating  wave. Under these conditions the mode becomes unstable and the flow
acts at the resonant layer as an energy source. In most of the aforementioned
studies it has been assumed that the wavelength is shorter or similar to the
tube radius. This is not the case for standing kink oscillations in coronal
loops which are precisely in the opposite regime, i.e., where the thin tube (TT)
approximation is applicable.

In the present paper we extend the previous studies about resonant absorption in
the presence of flow to the situation where the TT approximation is valid. We
start by reviewing the properties of kink MHD waves in a homogeneous tube with
an axial flow and study the nature of the waves, that change from being trapped
to leaky and eventually become KH-unstable. Then we consider a non-uniform tube
whose oscillation is damped by resonant absorption and  investigate how the
damping time is modified by the flow. In the following analysis three different
approaches are implemented. Firstly, we use existing theoretical work, mainly by
\citet{goossensetal92}, to calculate the changes in the period and damping rates
induced by the longitudinal flow using the thin tube thin boundary approximation
(TTTB). Secondly, under the TTTB assumption we derive a linear analytical
approximation to the damping rate and thirdly, solve the full resistive
eigenvalue problem without the TTTB assumption. Reassuringly, we find that the
three methods lead to essentially the same results.

\section{Basic features: uniform tube}  

We consider what we can call the {\it standard loop model}, a cylindrical
axi-symmetric flux tube of radius $R$ with a constant axial magnetic field $B_0$
and with a density contrast of $\rho_i/\rho_e$ where the indices ``i" and ``e"
describe quantities inside and outside the loop, respectively. Inside the loop
there is an axial flow denoted by $v_i$. For simplicity we assume that there is
no flow outside the tube so that  $v_e=0$. We start by recalling the analytical
results obtained for a uniform loop (no transition layer) in the $\beta=0$ case.
It is well-known that the effect of the flow introduces a shift in the frequency
of waves and that the known expressions for the dispersion relation without flow
can be used by simply replacing the frequency $\omega$ by its Doppler-shifted
counterpart, $\Omega=\omega-k v$, $k$ being the wavenumber along the tube. The
dispersion relation of MHD waves was derived in \citet{goossensetal92}
\citep[see also][]{terrahometal03, soleretal08}. In the TT approximation ($k R
\ll 1$) \citet{goossensetal92} found that the frequency of the kink MHD wave
modified by the flow, $\omega_{kf}$, (see their equation Eq.~[83]) is given by

\begin{equation}\label{freq} \omega_{kf} = k \frac{\rho_i
v_i}{\rho_i + \rho_e}\pm \omega_{cm}, \end{equation} where $\omega_{kf}$ is the
frequency of the kink wave modified by the flow, and  $\omega_{cm}$ is  the
frequency in the center-of-mass frame as in \citet{holletal90},
\begin{equation}\label{freqcm} \omega_{cm} = \left
\{\omega_k^2-\frac{\rho_i\rho_e}{(\rho_i+\rho_e)^2}k^2 v_i^2\right\} ^{1/2}.
\end{equation} In the above equation $\omega_k$ is the classic kink frequency of a
thin static tube and is given by \begin{equation}\label{freqkink} \omega_k =
\left \{\frac{\rho_i \omega_{A,i}^2 + \rho_e \omega_{A,e}^2}{\rho_i + \rho_e}
\right \}^{1/2}, \end{equation} with $\omega_{A}$ being the local Alfv\'{e}n
frequency, defined as \begin{equation}\label{freqAlven} \omega_{A} = k
v_A. \end{equation}

The frequency $\omega_{kf}$ of the MHD waves is degenerate with respect to the
azimuthal wave number in the TT approximation. Our interest is in the kink waves
with azimuthal number $m=1$ since these waves are the ones that move the axis of
symmetry and the loop as a whole. The plus and minus sign in front of the second
term of Eq.~(\ref{freq}) represent two different waves, one propagating in the
same direction of the flow (forward) and the other propagating against the flow
(backward). The presence of the flow breaks the degeneracy of positive and
negative frequencies present when $v_i = v_e=0$.

\begin{figure}[!ht]\center{
\resizebox{8.25cm}{!}{\includegraphics{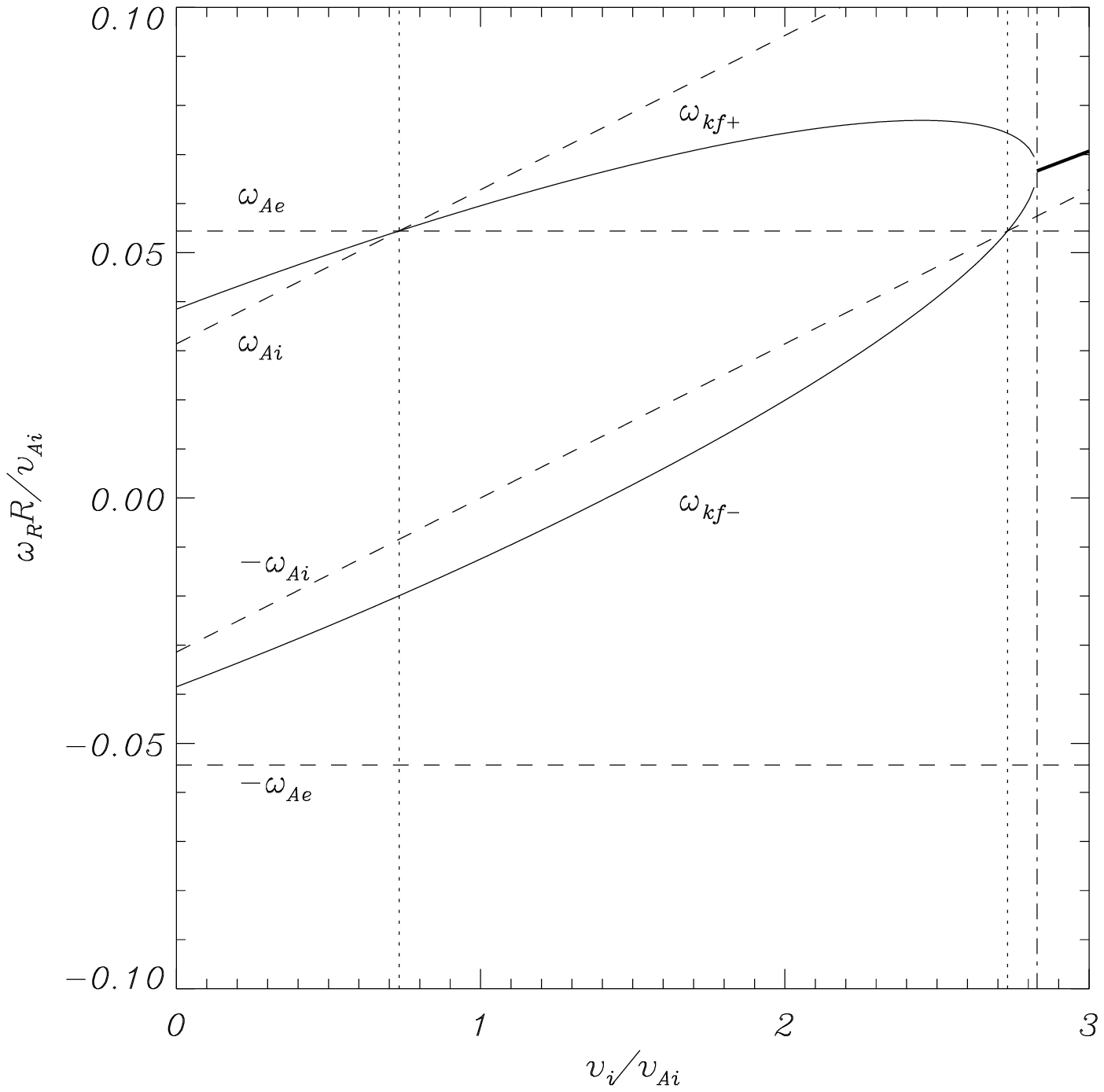}}{\bf
a)}\hspace{-0.0cm}  
\resizebox{8.25cm}{!}{\includegraphics{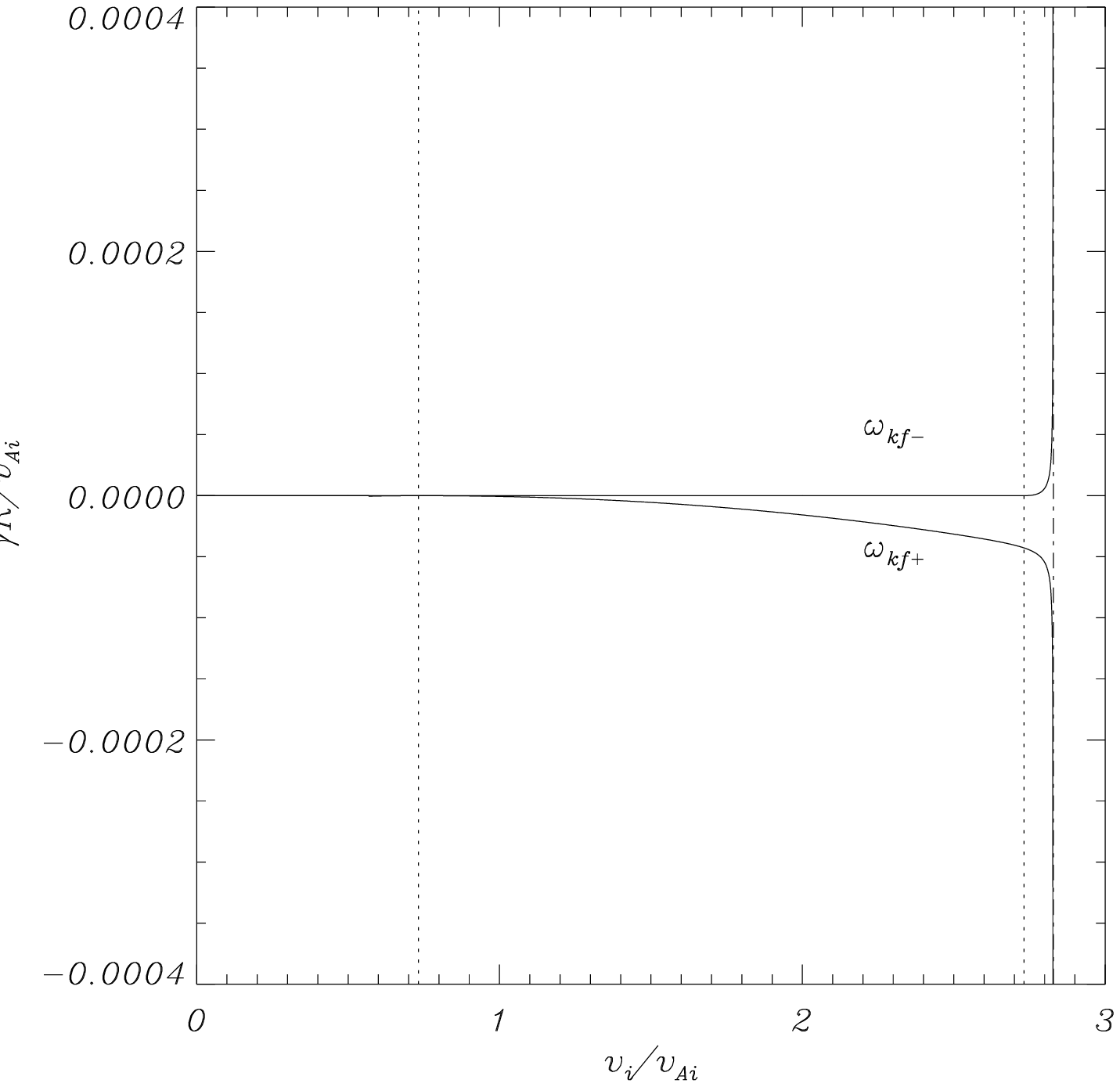}}{\bf
b)}\hspace{-0.0cm}}
\caption{\small {\bf (a)} Real and {\bf (b)} imaginary part of the frequency of
the forward ($\omega_{kf+}$) and backward ($\omega_{kf-}$) propagating waves as a
function of the internal flow ($v_i$) in units of the internal Alfv\'en speed. The curves intersect at the KH critical
flow, see dot-dashed line, given by Eq.~(\ref{khi}). The forward wave becomes
leaky before the KHI criteria, since the frequency of the mode crosses the
external Alfv\'en frequency ($\omega_{Ae}$), see dotted lines, at the value of the flow given
by Eq.~(\ref{leaky}) (in the TT approximation). The dashed lines represent  the Doppler
shifted Alfv\'en frequencies. In this plot $L=100R$,
$\rho_i/\rho_e=3$.}\label{khidiagram} \end{figure}

The second term of Eq.~(\ref{freq}) contains the condition for the
Kelvin-Helmholtz instability which occurs for any velocity shear in absence of a
magnetic field. When magnetic fields are present it is straightforward to see
that the square root in Eq.~(\ref{freq}) is negative when the flow is faster
than the critical value given by
\begin{equation}\label{khi}
\frac{v_i}{v_{Ai}}\geq\sqrt{2\left(1+\frac{\rho_i}{\rho_e}\right)}.
\end{equation}
The above condition means that fast flows compared to the internal
Alfv\'en speed are required for the Kelvin-Helmholtz instability to
occur \citep[see also][]{chandra61,ferrarietal81}. An equivalent problem has been studied in the context of propagating
transverse waves in coronal jets by \citet{farahanietal09}. These authors have
also found that in the observationally determined range of parameters, the waves
do not undergo either to the KHI or the negative energy wave
instability.

However, before the Kelvin-Helmholtz instability occurs, it may happen that the
frequency of the modes is above the external cut-off frequency ($\omega_{A,e}=k
v_{A,e}$), meaning that the wave becomes leaky.  The forward  propagating wave
becomes leaky when the following condition is satisfied

\begin{equation}\label{leaky}
\frac{v_i}{v_{A,i}}\geq\sqrt{\frac{\rho_i}{\rho_e}}- 1.
\end{equation}

Similar to the KH-instability, fast flows are required to generate a leaky wave.
Contrary to the static situation studied by \citet{goossensetal09}, in the
presence of flow an underdense loop is not required to have leaky modes when
$\beta=0$. Solving the dispersion relation in the TT approximation, the damping
of waves due to leakage is derived from the imaginary part of the frequency given by
\begin{equation}\label{gammarad}
\gamma=-\frac{\pi}{4} (k_e
R)^2\frac{\rho_e(\omega_{kf+}^2-\omega_{A,e}^2)}{(\rho_i+\rho_e)\, \omega_{cm}}.
\end{equation}
The decrement is proportional to the  square of $k R$ ($k_e\sim k$),
meaning that leakage is basically unimportant in the TT
approximation since the damping time ($\tau_D=1/\gamma$) is very large. Thus,
the damping due to MHD radiation in the presence of flow is a very inefficient
damping mechanism for the transverse loop oscillations, even in the presence of
fast flows.

An example of the dependence of the real ($\omega_R$) and imaginary ($\gamma$)
part of the frequency of the modes (fundamental forward and backward waves) on
the flow is shown in Fig.~\ref{khidiagram} for a particular equilibrium
configuration ($L=100R$, $\rho_i/\rho_e=3$). The domains where waves become
leaky or when a KHI occurs are clearly shown. When the frequencies of the two
modes merge for increasing velocity shear ($v_i$) the system becomes unstable.
However, note that the forward wave is always leaky before the system
becomes KH unstable (compare also Eq.~[\ref{leaky}] with Eq.~[\ref{khi}]). It is
important to mention here that we have not considered here the Principal Leaky
Mode which is a very peculiar solution of the dispersion relation
\citep[see][]{cally86,cally03} and instead have focused on the modes than are trapped
for a the static background.

Once we know the main effects of the flow on the kink MHD waves we need to know
in which regime of Fig.~\ref{khidiagram} we can match,  for example, the
observed standing kink oscillations.  The observations of flows in coronal loops
indicate that they are slow, therefore hereafter we focus on sub-Alfv\'{e}nic
flows rather than the super-Alfv\'{e}nic flows that might cause leakage and KH
instabilities. We concentrate on the regime $v_i/v_{Ai}<0.1$, thus according to
the previous analysis both the forward and backward waves are always trapped.
 This also prevents the presence of resonant flow instabilities which occur
when the frequency of the forward propagating wave shifts into the Doppler
shifted continuum of the backward propagating  wave.

\section{Waves in a Non-uniform tube}

Now let us consider a tube with a smooth variation of density and flow across
the loop cross-section. In particular we consider the case when $\rho$ varies
from its internal value $\rho_i$ to its external value $\rho_e$ in the interval
$[R-l/2,R+l/2]$ and the velocity changes from $v_i$ to $0$ in the interval
$[R-l^\star/2,R+l^{\star}/2]$. Under such conditions the process of resonant
absorption takes place and kink oscillations in coronal loops will damp
efficiently. The reader is referred to \citet{goossens08} and references therein
for a detail review on this kink wave damping mechanism.

 As in the previous Section we concentrate on propagating waves, the possible
excitation of standing waves in the presence of flow is discussed later.

\subsection{The TTTB approximation}\label{tttb}

In a non-uniform tube the imaginary part of the frequency (of the trapped

propagating  modes) is different from zero due to mode conversion at the inhomogeneous
layer ($\omega=\omega_R+i \gamma$). Some time ago \citet{goossensetal92}
derived an expression for the damping rate in the thin tube and thin boundary
($l\ll R$) approximation for incompressible MHD waves. For compressible waves
in a magnetic cylinder, using the loop model considered here, we obtain exactly
the same expression, given by (see their Eq.~[76])
\begin{equation}\label{gamma}
\gamma=-\frac{\rho_i^2(\Omega_i^2-\omega_{A,i}^2)^2}{2(\rho_i+\rho_e)\,
\omega_{cm}} \frac{1}{\rho(r_A)|\Delta|}\frac{|m| \pi}{r_A}. \end{equation}
As
usual $r_A$ is the resonant position that is calculated from
\begin{equation}\label{Omega} \Omega^2(r_A)=\omega_A^2(r_A), \end{equation} 
i.e. where there is a match between the Doppler shifted frequency and the local
Alfv\'en frequency. It is assumed that the real part of the frequency of the
resonantly damped mode is given by Eq.~(\ref{freq}). From a physical point of
view, the condition given by Eq.~(\ref{Omega}) means that the eigenmodes
resonantly interact with the Alfv\'en continuum, which is Doppler shifted as a
result of flow. 

The factor $\Delta$ in the denominator of Eq.~(\ref{gamma}) is
\begin{equation}\label{Delta}
\Delta=\frac{d}{dr}\left(\Omega^2(r)-\omega_A^2(r)\right),
\end{equation}
which contains a term with the derivative of the flow in the radial direction, 
absent in the static situation, that can increase or decrease the value of
$\Delta$.

Given a particular density and velocity profile, the different variables in
Eq.~(\ref{gamma}) can be evaluated. For simplicity, we use the
well known sinusoidal profile for the density given by
\begin{eqnarray}
\rho(r)=\left\{
\begin{array}{lll}
\rho_{i},&  0\leq r< R-l/2, \\
\frac{\rho_i}{2}\left[\left(1+\frac{\rho_e}{\rho_i}\right)
-\left(1-\frac{\rho_e}{\rho_i}\right)\sin\frac{\pi\left(r-R\right)}{l}\right],&  R-l/2\leq r \leq R+l/2, \\
\rho_{e},& r> R+l/2.
\end{array}
\right.
\label{dens}
\end{eqnarray}

This convenient profile has been used in many studies about resonant absorption
\citep[e.g.][]{rudrob02,vandetal04,arretal05,terretal06b}. In order to make the
mathematical approach more tractable we also assume that the variation of the flow
speed is sinusoidal inside the loop layer, i.e.,
\begin{eqnarray}
v(r)=\left\{
\begin{array}{lll}
v_{i}, & \; 0\leq r< R - l^{\star}/2, \\
\frac{v_i}{2}\left[1-\sin{\frac{\pi(r-R)}{l^{\star}}}\right],
& \; R-l^{\star}/2\leq r \leq R+l^{\star}/2, \\
0, & \; r> R+l^{\star}/2.
\end{array}
\right.\label{vel}
\end{eqnarray}
The flow is variable over a layer of thickness $l^{\star}$ which is not
necessarily equal to the characteristic thickness of the layer $l$ where the
density is  non-constant. In general, in preceding studies it has been assumed a
discontinuous flow (at the boundary of the loop), except by
\citet{erdtaro03a,erdtaro03} who considered a linear profile to model MHD waves
in the Earth's magnetotail.

\begin{figure}[!ht] 
\includegraphics[width=9.25cm]{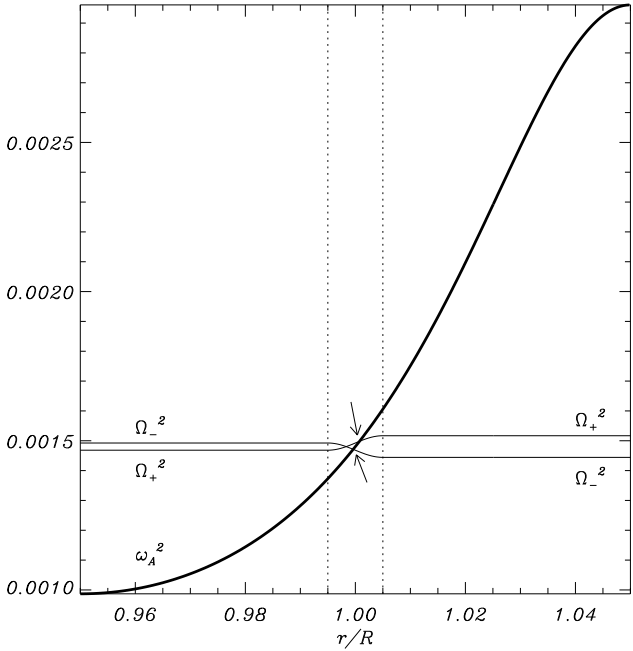} \caption{Forward
($\Omega_+$), backward ($\Omega_-$) and local Alfv\'en frequency (thick line) as a function of
the radial position. The arrows indicate the location of the resonances. In this plot $v_i/v_{Ai}=0.02$, $l/R=0.1$, $l^\star/R=0.01$, $L=100R$ and
$\rho_i/\rho_e=3$. The dotted vertical lines mark the limits of the inhomogeneity of the
flow.}\label{Omegawa1}
\end{figure}

For the profiles given by Eqs.~(\ref{dens}) and (\ref{vel}) it turns out that
Eq.~(\ref{Omega}) is a transcendental equation for the resonant position $r_A$.
This equation is solved using standard numerical techniques. Depending on the
spatial scales of the density and velocity we distinguish two different regimes,
$l^{\star}\gtrsim l$ and the asymptotic case $l^{\star} \ll l$.

The analysis of the first situation is rather simple  \citep[see
also][]{peredotata90}, since the the forward propagating wave has always a single
resonant position in the range $R<r_A<R+l/2$, while the resonant position of the
backward wave is situated in the range $R-l/2<r_A<R$. This behaviour is easily
understood from Fig.~\ref{Omegawa1}, where we have plotted the Doppler shifted
frequencies and the Alfv\'en frequency as a function of the radial coordinate.
The resonant positions are located at the intersection of $\Omega^2$ with
$\omega_A^2$ (see arrows). Note that Fig.~\ref{Omegawa1} also shows that if the
Alfv\'en frequency is discontinuous (jump in density, $l=0$) there are no
resonances (implying no damping) since $\Omega^2$ will never intersect the curve
corresponding to $\omega_A^2$.

Once the resonant position $r_A$ is determined $|\Delta|_{r_A}$ is evaluated
and we finally obtain the value of the damping rate $\gamma$ (using
Eq.~[\ref{gamma}]). A useful quantity that we can calculate is the the damping per period, given by 
\begin{equation}
\frac{\tau_D}{P}=\frac{|\omega_{kf}|}{|\gamma|}\frac{1}{2 \pi}.
\end{equation}

\begin{figure}[!ht] \includegraphics[width=9.25cm]{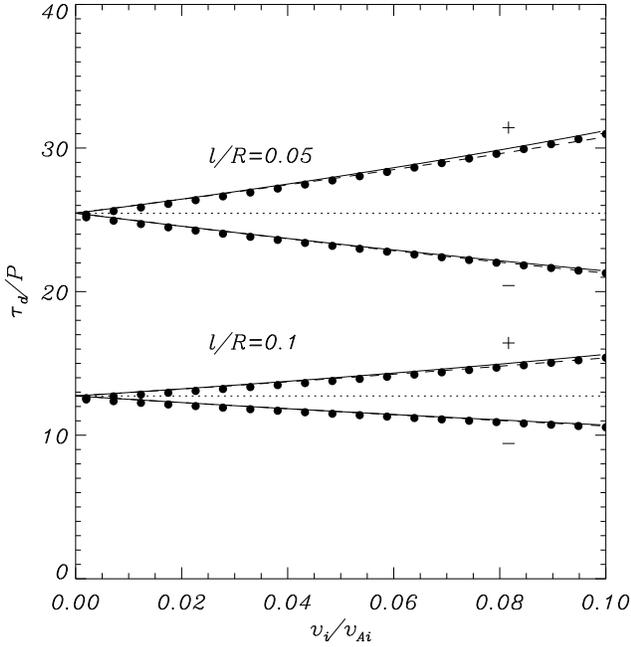} \caption{Damping
per period as a function of the flow inside the loop for the forward (+) and
backward (-) propagating waves. The solid lines represent the analytical
results calculated using Eq.~(\ref{freq}) and Eq.~(\ref{gamma}). The dashed
lines are the approximations of the damping per period using
Eq.~(\ref{L-omegakf}) and Eq.~(\ref{gammacomp}). The dots represent the full
numerical solution of the resistive eigenvalue problem. The horizontal dotted
lines are the damping per period in the static situation. For the curves with
$l/R=0.05, 0.1$ we have used $l^\star/R=0.05, 0.1$.} \label{tdoPan}
\end{figure}

In this expression we use the real part of the frequency given by
Eq.~(\ref{freq}). In Fig.~\ref{tdoPan} (see solid lines) $\tau_D/P$  is
represented for two different values of the characteristic widths of the layers,
$l/R=l^\star/R=0.05, 0.1$, as a function of the internal flow (recall that both
the frequency and the damping rate depend on the flow). The curves with positive
slope correspond to the forward waves while the ones with the negative slope
represent the backward wave. Figure~\ref{tdoPan} also shows that increasing the
strength of the flow increases the damping per period for the forward wave and
decreases it for the backward wave. This is in agreement with the results
of \citet{peredotata90}, the shifts of the frequency modify the location of the
resonant surfaces in such a way that one of the natural modes is closer to the
resonance while the other is further away from the resonance relative to the
static situation. As a consequence, one mode is damped more efficiently
than the other. Nevertheless, the effect of the flow does not significantly
change the damping per period for the regime considered here
($0<v_i/v_{Ai}<0.1$). Hence the mechanism of resonant absorption is very robust
in the presence of an internal flow. The thicker the layer, the faster the
attenuation (c.f., the results for $l/R=0.1$ with the results for $l/R=0.05$), a
result already known for the static case. We also see that the change of
$\tau_D/P$ is quite smooth with respect to $v_i$.

\begin{figure}[!ht]
\includegraphics[width=9.25cm]{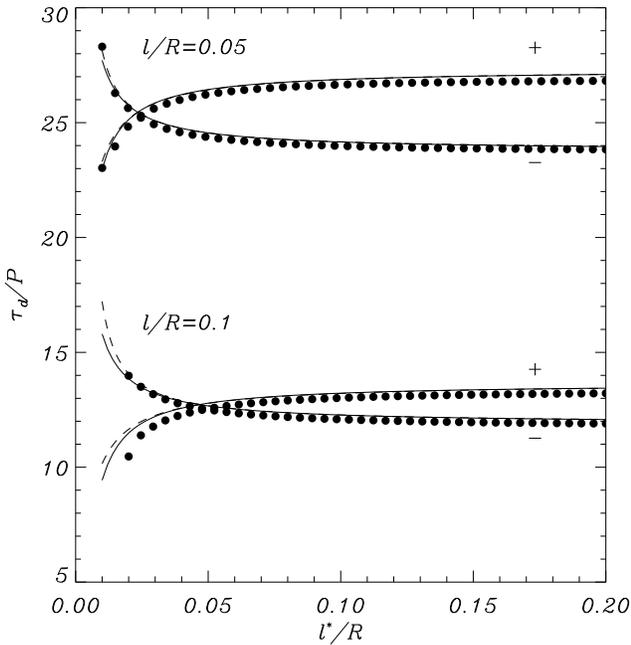}
\caption{
Damping per period for the forward and
backward propagating waves as a function of the width of the flow profile,
$l^\star$. We use the same notation as in Fig.~\ref{tdoPan}. In this plot 
$v_i/v_{Ai}=0.02$.}\label{tdoPlv}
\end{figure}

In Fig.~\ref{tdoPlv} we have represented the damping per period as a function of
$l^\star$ in units of loop radius for two different values of $l/R$. In this
plot, $l$, the characteristic scale of the density transition, is fixed (recall
we are still in the regime with $l^{\star}\gtrsim l$). For large values of
$l^\star$ compared to $l$ we see that the dependence is quite weak with the
thickness of the flow profile. The forward propagating wave has a larger damping
per period than the backward propagating wave. However, when $l^\star \lesssim
l$ the situation is reversed. The curves cross and the forward propagating wave
is attenuated  faster than the backward propagating wave, indicating that we are
at the threshold of a different regime.

\begin{figure}[!ht] 
\includegraphics[width=9.25cm]{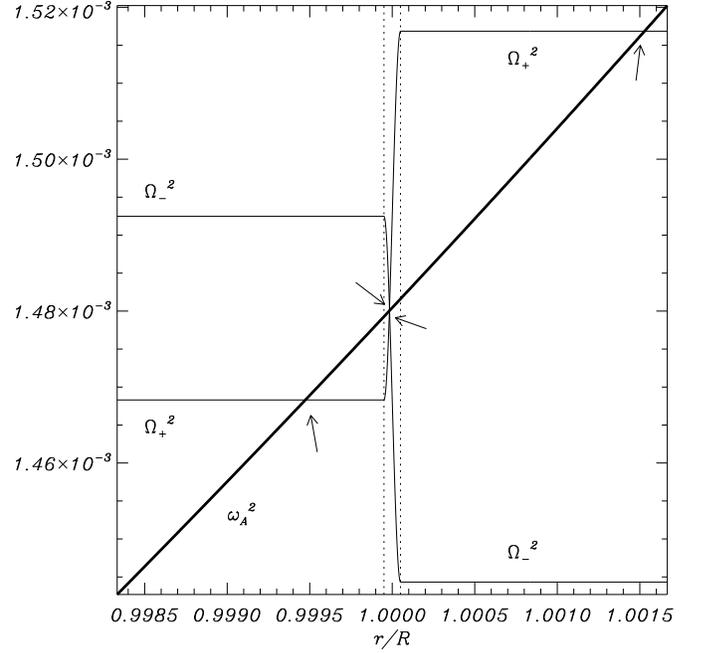} \caption{Forward
($\Omega_+$), backward ($\Omega_-$) and local Alfv\'en frequency (see thick
line) as a function of
the radial position. The arrows indicate the location of the resonances. In this plot $v_i/v_{Ai}=0.02$, $l/R=0.1$ and $l^\star/R=0.0001$, $L=100R$ and
$\rho_i/\rho_e=3$. The vertical dotted lines mark the limits of the inhomogeneity of the
flow.}\label{Omegawa2}
\end{figure}

Now let us concentrate on the situation when  $l^{\star}\ll l$, i.e., we
investigate the case with a very steep profile for the equilibrium flow
velocity. This is different to the regime discussed earlier ($l^{\star}\gtrsim
l$) in several aspects. In Fig.~\ref{Omegawa2} we have plotted a typical
example. It is easy to see that the forward wave can have now  three different
resonant positions  (see arrows). Apart from the resonance inside the
inhomogeneous velocity layer, the forward wave has two additional resonances,
one at $R-l/2<r_A<R-l^{\star}/2$ and the second at $R+l^{\star}/2<r_A<R+l/2$.
The backward wave still has a single resonance situated, as before, inside the
velocity layer. Although the situation is more complicated, we can still
understand the role of the resonances. The derivatives of $\Omega^2$ with
respect to $r$ at the resonance inside the velocity layer, located around $r=R$
(for both the forward and backward waves), become very large (in absolute value)
for small $l^{\star}$, thus dominating over the derivative of $\omega_A^2$ (see
Eq.~[\ref{Delta}]). This means that the factor $|\Delta|_{r_A}$ is large and
tending to infinity (for $l^{\star} \ll l$), therefore $\gamma$ will tend to
zero, i.e., they will not produce any damping. However, the other two resonances
of the forward wave still behave as ordinary resonances since the derivative of
the flow is zero (they are located outside the velocity layer where the flow is
constant) and in this situation the total damping of the mode will be finite due
to the combined contribution of the two resonances.

\subsection{Linear approximation of frequency and damping
time in the TTTB approximation}\label{lineartttb}

Visual inspection of Fig.~\ref{tdoPan} shows that the damping per period varies
smoothly and appears, to very good approximation, to be a linear
function of $v_i/v_{A,i}$. This has motivated us to derive
a linear approximation of the frequency $\omega_{kf}$ and the
damping rate $\gamma$ as function of $v_i/v_{A,i}$. Actually, it
turns out that
\begin{equation}\label{X}
x = \frac{k v_i}{\omega_k},
\end{equation}
is a more convenient variable for obtaining the linear approximation. Since
\begin{equation}\label{X1}
x = \frac{1}{\sqrt{2}} \left(\frac{\rho_i + \rho_e}{\rho_i}
\right)^{1/2} \frac{v_i}{v_{A,i}},
\end{equation}
it follows that $v_i/v_{A,i} \ll 1 $ is equivalent to $x \ll 1$.

The first order approximation to $\omega_{cm}$ (given by Eq.~[\ref{freqcm}]) is
\begin{equation}
\omega_{cm} = \omega_k,
\end{equation}
so that the linear approximation to $\omega_{k+}$ (for the forward wave) is
\begin{equation}\label{L-omegakf}
\omega_{kf+} = \omega_k \left( 1 + x \frac{\rho_i}{\rho_i + \rho_e}\right).
\end{equation}

In order to derive a linear approximation to $\gamma$ we need some intermediate
results to be used in Eq.~(\ref{gamma}). The linear approximation to
quantity $\rho_i (\Omega_i^2 - \omega_{A,i}^2)^2$ is
\begin{eqnarray}
\rho_i \left(\Omega_i^2 - \omega_{A,i}^2\right)^2 &=& \frac{
\rho_i^2 \rho_e^2}{\left(\rho_i + \rho_e\right)^2} \left (
\omega_{A,e}^2 - \omega_{A,i}^2 \right )^2 \left \{1 - 4
\frac{\omega_k^2}{\omega_{A,e}^2 -
\omega_{A,i}^2} x \right \},\nonumber \\
\end{eqnarray}
and the linear approximation to $\rho(r_A) \Delta $ can be written as
\begin{equation}\label{rdelt}
\rho(r_A) \Delta = \rho(r_A)\, \omega_k^2 \left \{\frac{
x}{ l_v} - \frac{\left(1 + y\right)}{ 
l_{\rho}} \right \}.
\end{equation}
Here $l_{\rho}$ and $l_v$ are the length scales of variation of
density $\rho$ and velocity $v_i$, respectively. They are defined as
\begin{equation}
\frac{1}{l_{\rho}} = \frac{1}{\rho(r_A)} \left| \frac{d
\rho}{d r}  \right|_{r_A},
\end{equation}
\begin{equation}
\frac{1}{l_{v}} = \frac{
1}{v(r_A)} \left| \frac{ d v}{d r}  \right|_{r_A}.
\end{equation}

Note that $l_{\rho}$ and $l_{v}$ are not equal to the width of the non-uniform
layer of density $l$ nor to the width of the non-uniform layer of velocity
$l^\star$. E.g, for the sinusoidal profile  $v(r_A)=v_i/2$ (since we assume that
$r_A=R$) and $\rho(r_A)=(\rho_i+\rho_e)/2$, the resultant characteristic spatial
scales are
\begin{equation}
l_{\rho} = \frac{ l}{ \pi}
\frac{ \rho_i + \rho_e}{ \rho_i - \rho_e},
\;\; l_v = \frac{ l^{\star}}{ \pi}.
\end{equation}
The quantity $y$ in Eq.~(\ref{rdelt}) is defined as
\begin{equation}
y = x\, \frac{\rho_i - \rho_e}{\rho_i +
\rho_e}.
\end{equation}
If we assume that $l_{\rho} \approx l_v$ then 
\begin{equation}
\frac{ x}{ l_v} \ll  \frac{ (1
+ y)}{l_\rho}.
\end{equation}
In this case the linear approximation to $ 1 /(\rho(r_A) \left| \Delta
\right|_{r_A})$ is
\begin{equation}
\frac{ 1}{ \rho(r_A) \left|\Delta \right |_{r_A}} =
\frac{ l_{\rho}}{ \rho(r_A)\, \omega_k^2} \left(1
- y + x J\right).
\end{equation}
Here $J$ is a factor which measures the relative importance
of the non-uniformity of the flow to that of density and it is defined
as
\begin{equation}
J = \frac{ l_{\rho}}{ l_v}.
\end{equation}
With all terms approximated linearly, the approximation of the
damping rate, $\gamma$, becomes
\begin{eqnarray}
\gamma=&-&
\frac{ \pi}{ 8}\frac{
l_{\rho}}{ \rho(r_A)}\frac{ (\rho_i -
\rho_e)^2}{ \rho_i - \rho_e} \, \omega_k \times \nonumber \\ 
&& \left \{ 1 + x \left [ -
\frac{ 8 \rho_i \rho_e}{ (\rho_i - \rho_e)
(\rho_i + \rho_e)} - \frac{ 2 \rho_i}{
\rho_i + \rho_e} + \frac{ 2 v(r_A)}{ v_i}
+ J \right] \right \}.\nonumber \\
&&
\end{eqnarray}
If we repeat the analysis for the backward wave we obtain the same expression with
a change in the sign in front of $x$. 

The final expression  for the damping rates of the two waves (forward and
backward propagating) using the sinusoidal profile of density and velocity
reduces to

\begin{eqnarray}\label{gammacomp}
\gamma= &-&
\frac{1}{4} \frac{l}{R} \frac{(\rho_i -
\rho_e)}{(\rho_i + \rho_e)} \omega_k \times \nonumber \\
&&\left \{ 1 \pm \frac{k v_i}{\omega_k} \left [- 
\frac{ 8 \rho_i \rho_e}{(\rho_i - \rho_e)
(\rho_i + \rho_e)} - \frac{ \rho_i -
\rho_e}{\rho_i + \rho_e} + \frac{(\rho_i + \rho_e)}
{(\rho_i - \rho_e)} \frac{l}{l^\star}\right] \right \}.\nonumber \\
&&
\end{eqnarray}

\noindent The $\pm$ sign corresponds to forward and backward propagating waves 
respectively. When the flow is zero, the second term inside the curly brackets
vanishes and we recover the formula of the damping in the static equilibrium
\citep[see for example][]{rudrob02}. In the non-static case since we have
assumed that $l_{\rho} \approx l_v$ ($l \approx l^\star$) the last two terms
inside the square brackets are of the same order and we see that the imaginary
part of the frequency linearly decreases with the flow for the forward
propagating wave and increases for the backward wave. Using the linear
approximation to the frequency (Eq.~[\ref{L-omegakf}]) and damping rate 
(Eq.~[\ref{gammacomp}]) it is straight forward to calculate the damping per
period. The results are represented in Fig.~\ref{tdoPan} (see dashed lines). We
see that the analytical approximations agree very well with the full solution
based on the calculation of $r_A$ and the evaluation of Eq.~(\ref{freq}) and
Eq.~(\ref{gamma}). Note that the dependence of the damping rate on the spatial
variation of the flow across the loop boundary, $l^\star$, is present in
Eq.~(\ref{gammacomp}). In Fig.~\ref{tdoPlv} we have plotted the results using
this expression (see dashed lines). Again, we find an excellent agreement
between the two curves for both the forward and backward waves.

For the regime $l^{\star}\ll l$ it is possible to derive useful information from
the linear approximation. Using the assumption $l_v\ll l_\rho$ it is easy to see
that the damping rate of the resonance inside the velocity layer (for both the
forward and backward waves) is proportional to $l^ \star$, which means, as we
have already anticipated in Section~\ref{tttb}, that the contribution of this
resonance to the total damping tends to zero (damping time tending to infinity)
for a purely discontinuous velocity profile. This is the behaviour already found
in Fig.~\ref{tdoPlv} for the backward wave. Moreover, we can estimate the total
damping of the two regular resonances of the forward wave (see
Fig.~\ref{Omegawa2}) by adding the individual damping rates. It turns out that
the total damping rate for the forward wave is simply
\begin{eqnarray}\label{gammacompfor} \gamma= &-& \frac{1}{2} \frac{l}{R}
\frac{(\rho_i - \rho_e)}{(\rho_i + \rho_e)} \omega_k \times \nonumber \\ &&\left
\{ 1 - \frac{k v_i}{\omega_k} \left [  \frac{ 8 \rho_i \rho_e}{(\rho_i - \rho_e)
(\rho_i + \rho_e)} + \frac{ \rho_i - \rho_e}{\rho_i + \rho_e} \right] \right \}.
\end{eqnarray} 

\noindent This is the asymptotic value for the forward wave when $l^ \star$
tends to zero. Note that this is twice the damping rate of the situation with a
very smooth velocity profile (see Eq.~[\ref{gammacomp}] when $l^ \star\gg l$)
indicating a more efficient attenuation (half the damping time).

\subsection{Beyond the TTTB approximation: full resistive eigenvalue problem}

The results of the previous Sections are based on the TT approximation. It is
known that without flows this approximation works very well even for thick
layers. However it remains to be confirmed whether this assumption is still
valid in the presence of flows. For this reason, we go beyond the TTTB
approximation. In this case we solve the full problem numerically. We follow the
approach of  \citet{terretal06b}. To study the quasi-mode properties, the
eigenvalue problem given by equations~(1)--(5) in  \citet{terretal06b}, plus the
additional terms due to the flow, is solved. A time dependence of the form $e^{i
\omega t}$ is assumed and the problem is solved numerically using a code based
on finite elements. As boundary conditions we impose that the velocity
components are zero for $r\rightarrow \infty$. In practice, the condition is
applied at $r=r_{max}$, and then it is necessary to check that the results do
not depend on this parameter. On the other hand, at $r=0$ it is imposed that
$\partial v_r/\partial r=0$, i.e., we select the regular solution at the origin,
while the rest of the variables are extrapolated. All the variables and the
eigenfrequency are assumed to be complex numbers, since we are interested in
resonantly damped modes. We include resistivity to avoid the singular behaviour
of the ideal MHD equations at the resonances. The resistive eigenvalue problem
is solved and we obtain the real and the imaginary part of the frequency which
must be independent of the magnetic Reynolds number that we use in the
computations \citep[][]{poedtskerner91}.

The results of the calculations for $l^{\star}\gtrsim l$ are plotted in
Fig.~\ref{tdoPan} (shown by dots). The agreement with the analytical
calculations, using the TTTB approximation, is remarkable. The numerical curves
almost overlap with the analytical ones. In Fig.~\ref{tdoPlv} (shown by dots) we
represent the damping per period as a function of $l^\star$ and find the same
behaviour as in the analytical expression. With these results we are even more
confident about the method used in Section~\ref{tttb} and about the analytical
expressions derived in Section~\ref{lineartttb}.

For the regime $l^{\star}\ll l$ the numerical method we are using fails since
the thinner the layer (in density or velocity) the larger the Reynolds number
required for the damping time to be independent of the dissipation. A method
based on the application of the jump conditions at the resonance or resonances,
used for example by \citet{tirryetal98} or  \citet{andriesetal00,andgooss01}, is
more appropriate but since this is not the main focus of this paper it will not
be further investigated here.

\subsection{The standing wave problem}

The results presented in the previous sections correspond to two propagating
waves, one propagating in the direction of the flow, $\omega_{f+}$, and the other
travelling in the opposite direction, $\omega_{f-}$. In general, an initial
perturbation will excite these two modes at the same time and the system will
oscillate with a combination of the two frequencies. If the frequencies are real
the superposition of the two modes (with the same amplitude and phase) will have
the following form
\begin{eqnarray}\label{freqbeat}  \cos(\omega_{f+}
t)+&\cos&(\omega_{f-}t)=\nonumber \\&2&
\cos\left(\frac{\omega_{f+} +|\omega_{f-}|}{2}t\right)
\cos\left(\frac{\omega_{f+} - |\omega_{f-}|}{2}t\right). \end{eqnarray}
This equation shows that the loop has an oscillation frequency given by
$\frac{\omega_{f+} + |\omega_{f-}|}{2}$ (recall that $\omega_{f-}$ is negative for
slow flows) modulated by an envelope that oscillates at the beating frequency
$\frac{\omega_{f+} - |\omega_{f-}|}{2}$. In the absence of flow we have that
$\omega_{f+}=-\omega_{f-}$, and from Eq.~(\ref{freqbeat}) we find the well known
pure standing wave solution of the static case. For slow flows (the regime
suggested by the observations), the frequencies of the forward and backward
waves hardly differ, and the beating frequency is very small (i.e., a very large
envelope), meaning that to all practical purposes the behaviour of the system is
equivalent to that of a standing wave. When fast flows are considered the motion
of the loop is much more complex and it is  dominated by the beating frequency.

It must be noted that fact that the amplitude of the oscillation is quickly
damped with time, as the observations indicate, might favour the formation of a
standing wave when the flow is present. In this situation, when the damping
times of the forward and backward waves are shorter than the beating period
($|\gamma|\gg \frac{\omega_{f+} - |\omega_{f-}|}{2}$) the envelope of the
signal is dominated by the attenuation due to resonant absorption rather than by
the modulation due to the beating.

\section{Conclusions and Discussion}

We have studied the effect of a longitudinal flow on propagating kink oscillations of a
coronal loop and their damping, and have shown, in agreement with previous
studies, that under typical coronal conditions a longitudinal flow, which is
highly sub-Alfv\'enic, is unable to produce KH-unstable modes. It was also
demonstrated that leaky modes are generated by fast flows that have velocities
comparable to the local Alfv\'en velocity. Since observations show that flows
are at most, 10\% of the Alfv\'en speed, this means that  forward and the
backward waves must always be trapped in coronal loops.  Moreover, the
forward wave never enters into the Doppler shifted continuum of the backward
propagating waves (see Fig.~\ref{khidiagram}a) and so there are no resonant flow
instabilities for slow flows. Although instabilities due to longitudinal flows
are unlikely to occur in coronal loops, other kinds of instabilities, for
example produced by the azimuthal shear of the kink mode are possible
\citep[see][]{terradasetal08,terradas09,ballaietal09}.

It was demonstrated that the resonant damping mechanism due to non-uniform
density and flow at the loop boundary is not significantly altered by the
presence of the flow as long as the scale of inhomogeneity of the flow is
similar or larger than the scale of inhomogeneity of the density. We derived 
simple expressions for the linear approximation to the frequency and damping
rate as a function of the flow, for forward and backward propagating waves in
the TTTB limit. These simple formulae are very accurate, since they agree very
well with the numerical calculations of the full resistive eigenvalue problem.
The analytical expressions will facilitate future seismological applications
\citep[along the lines of those proposed by][]{arreguietal07,goossetal08}, since
now the damping rate contains the velocity flow as an additional parameter.
Using these expressions we can estimate the differences with respect to the
static situation. For example, for a loop with flows of $v_i=0.1 v_{Ai}$ and a
thickness of the layer in density and velocity of $0.05R$, the period decreases
a $6\%$ and the damping time increases a $14\%$ for the forward wave, while for
the backward wave the period increases a  $6\%$ and the damping time decreases a
$11\%$ compared to the purely static equilibrium case.

A physically peculiar situation takes place when the flow has a sharp transition
at the loop boundary (in the limit of $l^\star\ll l$). The backward wave is
transformed into an undamped mode even in the presence of a non-uniform density
transition. Conversely, the forward wave is more efficiently damped due to the
introduction of two new resonances outside the velocity transition layer.

 Finally, we must remark that the problem studied in this paper is an
initial value problem where the wavenumber, $k$, is assumed to be ral, and we
solve for the complex frequency $\omega$. Nevertheless, a more convenient
description of certain coronal loop problems would require to study the boundary
value problem, where the frequency is prescribed and one solves for the complex
longitudinal wavenumber. This is will the subject of a future work.

\begin{acknowledgements} J.T. and M.G. acknowledge support from K.U.Leuven via
GOA/2009-009. J.T. acknowledges the funding provided under projects
AYA2006-07637 (Spanish Ministerio de Educaci\'on y Ciencia) and PCTIB2005GC3-03
(Conselleria d'Economia, Hisenda i Innovaci\'o of the Government of the Balearic
Islands). In addition, J.T. thanks Jesse Andries, Gary Verth and  Roberto
Soler for their useful suggestions that helped to improve the original
manuscript. The present research was initiated while I.B. was a guest at Dept.
of Physics, UIB (Spain). I.B. acknowledges the financial support and warm
hospitality of the Dept. of Physics, UIB. I.B. was supported by NFS Hungary
(OTKA, K67746) and The National University Research Council Romania
(CNCSIS-PN-II/531/2007). We are grateful as well to an anonymous referee
whose comments and suggestions helped us to improve the paper.

\end{acknowledgements}

\end{document}